\documentclass[twocolumn,showpacs,preprintnumbers,prd,superscriptaddress,nofootinbib]{revtex4}
\usepackage{amsmath}
\usepackage{amsfonts}
\usepackage{amssymb}
\usepackage{graphicx}
\usepackage[titletoc]{appendix}
\usepackage{color}
\usepackage{float}

\begin{document}

\title{Cosmographic bounds on the cosmological deceleration-acceleration transition redshift in $f(\mathcal{R})$ gravity}

\author{Salvatore Capozziello}
\affiliation{Dipartimento di Fisica, Universit\`a di Napoli ``Federico II", Via Cinthia, I-80126, Napoli, Italy.}
\affiliation{INFN Sez.\ di Napoli, Monte S.\ Angelo, Via Cinthia, I-80126, Napoli, Italy.}
\affiliation{Gran Sasso Science Institute (INFN), Viale F. Crispi, 7, I-67100, L'Aquila, Italy.}

\author{Omer Farooq}
\affiliation{Department of Physics, Kansas State University, 116, Cardwell Hall, Manhattan, KS, 66506, USA.}

\author{Orlando Luongo}
\affiliation{Dipartimento di Fisica, Universit\`a di Napoli ``Federico II", Via Cinthia, I-80126, Napoli, Italy.}
\affiliation{INFN Sez.\ di Napoli, Monte S.\ Angelo, Via Cinthia, I-80126, Napoli, Italy.}
\affiliation{Instituto de Ciencias Nucleares, Universidad Nacional Aut\'onoma de M\'exico, M\'exico City, DF 04510, Mexico.}

\author{Bharat Ratra}
\affiliation{Department of Physics, Kansas State University, 116, Cardwell Hall, Manhattan, KS, 66506, USA.}

\begin{abstract}
We examine the observational viability of a class of $f(\mathcal{R})$ gravity cosmological models. Particular attention is devoted to constraints from the recent observational determination of the redshift of the cosmological deceleration-acceleration transition. Making use of the fact that the Ricci scalar is a function of redshift $z$ in these models, $\mathcal {R=R}(z)$, and so is $f(z)$, we use cosmography to relate a $f(z)$ test function evaluated at higher $z$  to late-time cosmographic bounds. First, we consider a model independent procedure to build up  a numerical $f(z)$ by requiring that at $z=0$ the corresponding cosmological model reduces to standard $\Lambda$CDM. We then infer late-time observational constraints on $f(z)$ in terms of bounds on the Taylor expansion cosmographic coefficients. In doing so we parameterize possible departures from the standard $\Lambda$CDM model in terms of a two-parameter logarithmic correction. The physical meaning of the two parameters is also discussed in terms of the post Newtonian approximation. Second, we provide numerical estimates of the cosmographic series terms by using Type Ia supernova apparent magnitude data and Hubble parameter measurements. Finally, we use these estimates to bound the two parameters of the logarithmic correction. We find that the deceleration parameter in our model changes sign at a redshift consistent with what is observed.
\end{abstract}

\maketitle

\section{Introduction}

The inclusion of a  cosmological constant $\Lambda$ in Einstein's equations is arguably the simplest way to produce accelerated cosmological expansion. The corresponding cosmological model, namely the $\Lambda$CDM model \cite{Peebles84}, predicts a currently accelerating cosmological expansion and is in fairly good agreement with current observations \cite{Sami}.\footnote{In this model, the current cosmological energy budget is dominated by $\Lambda$, with cold dark matter (CDM) in second place, and baryons a distant third. The $\Lambda$CDM model assumes the simplest form of CDM, which might be in conflict with some observations on structure formation \cite{lambdaobs}.} However, the $\Lambda$CDM model has some puzzling features \cite{Sola}. The first puzzle is that both matter and $\Lambda$ energy densities are comparable in order of magnitude today. The second puzzle is the huge difference between the observed $\Lambda$ and the corresponding quantum field theory naively-computed value. Perhaps these puzzles mean that the standard $\Lambda$CDM model is only a limiting case of more complete, and less puzzling, cosmological model. In such a model the role played by $\Lambda$ in the $\Lambda$CDM model might be generalized to another, more complex but still unknown, substance, often dubbed (dynamical) dark energy \cite{darkenergy}.

The cosmological constant can be considered to be a fluid with equation of state $p_{\Lambda}=-\rho_{\Lambda}$ relating the time-independent energy density $\rho_{\Lambda}$ and pressure $p_{\Lambda}$. A fluid with equation of state $p_X=\omega_X\rho_X$ relating the $X$-fluid energy density $\rho_X$ and pressure $p_X$ is a very simple (but incomplete \cite{Podariu}) model of dynamical dark energy density. The $\phi$CDM model \cite{PeeblesandRatra},  where dynamical dark energy density is modeled by a scalar field $\phi$ with potential energy density $V(\phi)$ is a complete and consistent model. Many models have followed this one during the last quarter century, and an evolving dark energy fluid may be responsible for the late time acceleration, if the corresponding equation of state parameter, $\omega$, is within the interval $-1<\omega\simeq -0.75$, for redshift $z\ll1$ \cite{mico}. However, none of them has managed to satisfactorily clarify the physical origin of dark energy, thus a definite explanation of the accelerated cosmological expansion remains elusive.

A major shortcoming of the dark energy paradigm is the lack of more fundamental first principles (less phenomenological), motivation for dark energy \cite{notes}. Another possibility, is to, instead, ascribe the observed accelerating cosmological expansion to modified gravity, a generalization of general relativity \cite{coppa4}. This is the underlying philosophy of  modified theories of gravity, which extend general relativity by adding further curvature invariants to the Einstein-Hilbert action \cite{coppa45}. Much discussed extensions of standard cosmology arise when the Ricci scalar is replaced by an analytic function of curvature, $f(\mathcal{R})$ \cite{curvatura}. Here the action is
\begin{equation}\label{azione}
{\cal{A}} = \int{d^4x \sqrt{-g} \left [ f(\mathcal{R}) +
{\cal{L}}_{m}\right]}\,,
\end{equation}
where ${\cal{L}}_{m}$ is the matter Lagrangian, involving generally both baryons and cold dark matter and $g$ is the determinant of the metric tensor. By varying the action with respect to the metric tensor $g_{\mu\nu}$, one infers the  field  equations
\begin{eqnarray}\label{filedeqs}
\mathcal{R}_{\mu \nu}f^{'}(\mathcal{R})-\frac{1}{2}f(\mathcal{R})g_{\mu
\nu}-(\nabla_{\mu}\nabla_{\nu}-g_{\mu
\nu}\nabla_{\alpha}\nabla^{\alpha})f^{'}(\mathcal{R})\nonumber \\  =8\pi T_{\mu \nu}\,.
\end{eqnarray}
Here a prime denotes a derivative with respect to $\mathcal{R}$, $\mathcal{R}_{\mu \nu}$ is the Ricci tensor, $T_{\mu\nu}$ is the standard energy-momentum tensor and we have set the Newtonian gravitational constant and the speed of light to unity $G=c=1$. Dark energy can therefore be considered to be a geometrical fluid that adds to the conventional stress-energy tensor, if one want to interpret this equation in the context of general relativity \cite{geometric}. In doing so, the task of determining the dark energy equation of state is replaced by trying to understand which $f(\mathcal{R})$ better fits current data. It has been argued that viable candidates of $f(\mathcal{R})$ are those that reduce to the $\Lambda$CDM model at $z\ll1$ \cite{beng}. This guarantees fairly good agreement with present observations, permitting us to ease the experimental problems associated with wrong choices of $f(\mathcal{R})$.

In the $\Lambda$CDM model, and in dynamical dark energy models, the dark energy density has only recently come to dominate the cosmological energy budget and so accelerate the cosmological expansion. Earlier on dark matter dominated, resulting in decelerating cosmological expansion. Recent cosmological measurements have lead to the first believable estimate of the decelerating-accelerating transition redshift $z_{\mathrm da}$, \cite{ratra13}, of order 0.75. Thus there is now some observational support for the dark energy idea at redshifts approaching unity.  Therefore it is of interest to see whether such data are also consistent with $f(\mathcal{R})$ gravity models. In this work we determine observationally viable $f(\mathcal{R})$ models by assuming the cosmological principle and using cosmography to constrain parameters \cite{cosmography1}. From the Taylor series expansion of the scale factor $a(t)$, cosmography can be used to numerically bound late time measurable quantities, e.g., the acceleration parameter, the jerk parameter, the snap, and so forth, \cite{cosmography2}. Possible departures from the standard $\Lambda$CDM model could be determined through the use of cosmography, which represents a  tool to pick the most viable class of $f(\mathcal{R})$ models \cite{cosmography3}. To this end, cosmography allows us to relate the expanded quantities of interest, i.e. the Hubble rate, luminosity distances, magnitudes, and so forth, in terms of observables \cite{cosmography4}. In this paper, we show that a particular Hubble rate, derived from a viable class of $f(\mathcal R)$ models, predicts a transition from decelerated to accelerated cosmological expansion at a transition redshift which is in fairly good agreement with the cosmographic series. This is an extension of the standard $\Lambda$CDM model with a logarithmic term that mimics the effect of the dark energy as a smoothly varying function of the redshift $z$. The corresponding acceleration parameter changes sign around $z\sim 0.75$, in agreement with the recent measurement \cite{ratra13}. We also rewrite $f(\mathcal{R})$ as a function of $z$, as a series in the scale factor, i.e. $a\equiv(1+z)^{-1}$. This allows us to describe the curvature dark energy fluid in terms of the more practical redshift variable. In turn, we determine observational bounds on the cosmographic series and on the expanded $f(\mathcal R)$, by combining the most recent Union 2.1 supernova apparent magnitude compilation \cite{suzuky} and Hubble rate measurements in the interval $z\in[0,2.8]$ \cite{ratra13,ultima1}, through the use of Monte Carlo analyses using the the Metropolis algorithm \cite{metro}. We obtain our fits by using ROOT \cite{root} and BAT \cite{bat}.

Our paper is structured as follows: In Sec.\ II we set the initial conditions on cosmological observables, through the use of cosmography. These initial conditions are useful when determining constraints on $f(z)$ at low redshift. In Sec.\ III we  relate these initial conditions to the modified Friedmann equations. In Sec.\ IV we describe the corresponding cosmological model, inferred from numerically solving the modified Friedmann equations, with the numerical bounds inferred from cosmography. Furthermore, we describe the obtained transition redshift and we discuss the numerical results by comparing our model to $\Lambda$CDM. In Sec.\ V model parameters are computed by using supernovae apparent magnitude and Hubble parameter measurements.  In Sec.\ VI we provide a physical interpretation of the free parameters of the model, relating them to derivatives of the Ricci scalar at the present time. Finally, Sec.\ VII is devoted to our conclusions.

\section{Cosmography and $f(\mathcal{R})$ gravity}

In this section  we relate $f(\mathcal{R})$ gravity to the  cosmographic series of observables \cite{cosm}. Cosmography provides a way of determining constraints on $f(\mathcal{R})$ and its derivatives at small redshift by using the fact that the Ricci scalar is a function of the redshift $z$. In so doing one obtains the corresponding $f(z)$ function in terms of the cosmographic series. This procedure allows for an evaluation of the $f(z)$ derivatives at $z=0$ through observational bounds, and constitutes a  scheme where general relativity  represents a limiting case of a more general theory \cite{lymy}. Higher order curvature terms are therefore reinterpreted as a curvature dark energy fluid, responsible for possible departures from the standard $\Lambda$CDM model.

We start by expanding the scale factor $a(t)$ in a Taylor series around the present time $t_0$,
\begin{equation}\label{aexpans}
a(t)-1=\sum_{k=1}^{\infty}\frac{1}{k!}\frac{d^k a}{dt^k}\Big|_{t=t_0}(t-t_0)^k.
\end{equation}
The leading cosmographic series terms---the Hubble, deceleration, and jerk parameters---are
\begin{subequations} \label{eq:CScoeff}
\begin{align}
     H(t) &= \frac{1}{a}\frac{da}{dt}, \\
    q(t) &= -\frac{1}{a     H^2} \frac{d^2a}{dt^2}, \\
    j(t) &= \frac{1}{a     H^3} \frac{d^3a}{dt^3}.
\end{align}
\end{subequations}
We can use such quantities to study the kinematics of the universe \cite{turnercosmografia}, without  postulating a cosmological model. In this sense, cosmography is a model-independent technique that may be able to establish whether a given cosmological model is favored or not with respect to  other cosmological models.

On the other hand, this kinematical cosmographic series approach has more parameters that must be constrained by data than do the simplest cosmological models. To simplify the problem, we assume that space curvature is zero.\footnote{In the $\Lambda$CDM model, where the dark energy density is time independent, cosmic microwave background anisotropy measurements indicate the spatial curvature is at most very small \cite{Ade}. However, if the dark energy density varies with time, the observational bounds on space curvature are not so restrictive \cite{Anatoly2}.} We also assume the validity of the cosmological principle. These few assumptions simplify the problem and allow us to use kinematical cosmography as a model-independent tool to determine which of the various $f(\mathcal R)$ models are compatible with current observations. With more and better-quality near-future data, \cite{S.Podariu}, it should be possible to also constrain space curvature.

The cosmographic series terms we include in our analyses are the Hubble rate $H$, the acceleration parameter $q$, and the variation of acceleration $j$. Although additional coefficients may be added in the cosmographic analysis, the amount and the quality of current data requires that we limit ourselves to these three terms. These terms are sufficient to allow us to determine how the universe is currently speeding up and how the acceleration varies as the universe expands. The physical meaning of each term is as follows. The  Hubble rate is the first derivative with respect to the cosmic time of the logarithm of the scale factor $a$. The  acceleration parameter $q$ indicates how much the universe is currently accelerating. Taking space curvature to vanish, a currently accelerating universe has $-1 \leq q_0 \leq  0$ (where $q_0$ is the value of $q$ at the present time), the limit $q_0=-1$ represents a perfect de Sitter universe totally dominated by a cosmological constant, and $q(z_{\rm{da}})=0$ corresponds to the transition redshift between accelerating and decelerating expansion $z_{\rm{da}}$ \cite{at,att}. In turn, its variation $j$ should be positive today, so that $q$ changes sign as the universe expands. For the $\Lambda$CDM model, the jerk parameter $j_0=1$ at all times \cite{vs2}.

Expanding the luminosity distance\footnote{The luminosity distance $d_\mathcal L=\sqrt{\mathcal{L}/(4\pi l)}$ where $\mathcal L$ is the absolute luminosity and $l$ the apparent luminosity of the source.} $d_\mathcal L$ in a Taylor series in redshift $z$,
\begin{equation}\label{gt852369}
d_\mathcal L=  \sum_{n=1}^{\infty}\frac{1}{n!}\frac{d^nd_\mathcal L}{dz^n}\Big|_{z=0}z^n\,,
\end{equation}
and truncating to second order in $z$, yields
\begin{eqnarray}\label{dlinterminidiz}
d_\mathcal L \approx  &{\frac{z}{H_0}} {\Bigl[ 1 + \frac{z}{2} (1 - q_0) +
    \frac{z^2}{6} (3q_0^2 + q_0 - j_0 - 1  )\Bigr]}.
\end{eqnarray}
This provides a way to compare $d_\mathcal L$ to the observable cosmographic series. Indeed, one can fit the luminosity distance to cosmological data, and determine bounds on the cosmographic series, without  postulating a cosmological model to define $d_\mathcal L$.

Measuring the cosmographic parameters through the use of Eq.\ ($\ref{dlinterminidiz}$) has the disadvantage that the coefficients depend on combinations of $H_0,q_0$ and $j_0$. Indeed, we measure the ratios $(1-q_0)/{H_0}$ and $(3q_0^2+q_0-j_0-1)/{H_0}$, and not $H_0,q_0$ and $j_0$ independently. As a consequence of this the cosmographic coefficients degenerate and the corresponding errors can be large. The degeneracy can be alleviated by first measuring $H_0$ alone. One technique we adopt in this work consists of fitting $z\leq0.36$ data by assuming the first-order luminosity distance term $d_{\mathcal L} = z/H_0$. We then perform numerical fits using Eq.\ ($\ref{dlinterminidiz}$) with the obtained $H_0$ and determine the corresponding $q_0$ and $j_0$. This prescription may alleviate, in principle, the degeneracy  of $q_0$ and $j_0$ in terms of $H_0$. In the following sections, we perform a number of cosmological tests, allowing all the parameters to vary freely, by fixing $H_0$ by using \textit{Planck} data, and by fixing $H_0$ through the technique discussed above.


By using the definition of redshift in terms of the scale factor we have
\begin{equation}\label{zdit}
\frac{d\log(1+z)}{dt}=-{H}(z)\,,
\end{equation}
and we can rewrite the Ricci scalar $\mathcal R$ as a function of $z$,
\begin{equation}\label{eq: constr}
\mathcal{R} = 6 {H}\left[ (1+z)\ {H}_{z} - 2 {H}\right],
\end{equation}
where $H_z$ is the derivative of $H$ with respect to $z$.
It is straightforward to express the present epoch values of $\mathcal{R}$ and its derivatives in terms of $H_0$ and derivatives. We have
\begin{subequations}
\label{eq:CScoeff1}
\begin{align}
\label{Rinz0}
\mathcal{R}_0 =\, & 6{H}_0\left({H}_{z0}-2{H}_0\right)\,,\\
\mathcal{R}_{z0} =\,& 6{H}_{z0}^2-{H}_0(3{H}_{z0}-{H}_{2z0}),
\label{Rinz0bis}
\end{align}
\end{subequations}
where $H_{z0}$ and $H_{2z0}$ are the first and second $z$ derivative of $H$ evaluated at the present time.
Since
\begin{subequations}
\begin{align}
\label{eq:CSoftime}
q&=-\frac{1}{{H}^2}\frac{d{H}}{dt} -1\,, \\
j&=\frac{1}{{H}^3}\frac{d^2 {H}}{dt^2}-3q-2\,,
\end{align}
\end{subequations}
\noindent we have
\begin{subequations}
\label{Hpunton}
\begin{align}
\label{Hpunto}
\frac{d{H}}{dt} =& -{H}^2 (1 + q)\,,\\
\frac{d^2H}{dt^2} =& {H}^3 (j + 3q + 2)\,.\label{Hpuntone}
\end{align}
\end{subequations}
Using Eq. (\ref{zdit}), we can rewrite Eqs. (\ref{Hpunton}) in terms of the cosmographic series terms only, obtaining
\begin{subequations}
\label{Hinz0bi}
\begin{align}
{H}_{z0}& = {H}_0(1+q_0)\,,\label{Hinz0}\\
{H}_{2z0}& = {H}_0(j_0-q_0^2)\label{Hinz0bis}\,.
\end{align}
\end{subequations}

\noindent Then, using Eqs. (\ref{eq:CScoeff1}) and (\ref{Hinz0bi}), we can express $\mathcal{R}$ and its derivatives as functions of $ H_0,\ q_0$, and $j_0$. This represents the first prescription for rewriting the $f(z)$ derivatives in terms of cosmographic parameters.

\section{Modified Friedmann Equations in $f(\mathcal{R})$ gravity}

We now discuss how the cosmological Friedmann equations for general relativity are modified in $f(\mathcal{R})$ gravity  and how these modifications can be viewed as ``dark energy" contributions to the Friedmann equations of general relativity. In addition, we describe how the derivatives of $f(z)$ can be related to the cosmographic series.

In the case of pressureless matter ($p_\mathrm{m}=0$), the modified Friedmann equations in $f(\mathcal{R})$ gravity are
\begin{subequations}
\label{primieroboth}
\begin{align}
&{H}^2 = \frac{1}{3} \left [ \rho_{\mathrm{curv}} + \frac{\rho_\mathrm{m}}{f'(\mathcal{R})} \right
]\,,\label{primiero}\\
&2 \dot{{H}} + 3{H}^2= - p_{\mathrm{curv}}\,,\label{secundo}
\end{align}
\end{subequations}
\noindent with nonrelativistic matter density $\rho_{\mathrm{m}}\propto a^{-3}$. Here the overdot denotes a time derivative and the prime represents a derivative with respect to the curvature $\mathcal R$. The curvature corrections can be used to describe a dark energy fluid, responsible for the current cosmological acceleration. These are the energy density
\begin{equation}\label{eq:rhocurv}
\rho_{\mathrm{curv}} = \frac{1}{f'(\mathcal{R})} \left \{ \frac{1}{2} \left[ f(\mathcal{R})  - \mathcal{R}
f'(\mathcal{R}) \right] - 3 {H} \dot{\mathcal{R}} f''(\mathcal{R}) \right \} \,,
\end{equation}
and the pressure obeying the equation of state $p_{\mathrm{curv}}=\omega_{\mathrm{curv}}\rho_{\mathrm{curv}}$ with equation of state parameter
\begin{equation}
\omega_{\mathrm{curv}} = -1 + \frac{\ddot{\mathcal{R}} f''(\mathcal{R}) + \dot{\mathcal{R}} [ \dot{\mathcal{R}}
f'''(\mathcal{R}) - {H} f''(\mathcal{R})]} {\left [ f(\mathcal{R}) - \mathcal{R} f'(\mathcal{R}) \right ]/2 - 3
{H} \dot{\mathcal{R}} f''(\mathcal{R})}\,. \label{eq: wcurv}
\end{equation}

It is convenient for our purposes to work in terms of $f(z)$, i.e.\ $f(\mathcal{R})$ as a function of redshift $z$. Since $\mathcal{R}=\mathcal{R}(z)$, we have
\begin{subequations}
\begin{align}\label{zuzzu}
f'(\mathcal{R}) =\, & \mathcal{R}_z^{-1}f_z\,,\\
f''(\mathcal{R})=\, & (f_{2z}\mathcal{R}_z - f_z\mathcal{R}_{2z})\mathcal{R}_z^{-3}\,,\\
f'''(\mathcal{R})=\, &\frac{f_{3z}}{\mathcal{R}_z^3} - \frac{f_z\, \mathcal{R}_{3z}+3f_{2z}\,
\mathcal{R}_{2z}}{\mathcal{R}_z^4}+\frac{3f_z\, \mathcal{R}_{2z}^2}{\mathcal{R}_z^5}\,,
\end{align}
\end{subequations}
which relate $f(\mathcal R)$ to $f(z)$ and derivatives. To evaluate the derivatives of $f(\mathcal R)$ in terms of the Hubble rate, we make use of the following identities
\begin{subequations}\label{rdotrddot}
\begin{align}
\dot {\mathcal{R}}=&-(1+z){H}\mathcal{R}_z\,,\\
\ddot {\mathcal{R}}=&(1+z)H\big[H\mathcal{R}_z+(1+z)(H_z\mathcal{R}_z+{H}\mathcal{R}_{2z})\big]\,.
\end{align}
\end{subequations}

We consider $f(\mathcal{R})$ gravity models which are consistent with the Solar System tests \cite{salo}. Thus the gravitational constant does not depart from its observed value. Keeping in mind such prescriptions, we fix the initial conditions on $f(z)$ and its derivatives, by relating $f(z)$ and $f^\prime(z)$ to the cosmographic series through
\begin{subequations}\label{f0fz0fzz0dopo}
\label{etunoboth}
\begin{align}
f_0=\,&2{H}_0^2(q_0-2),\label{etuno} \\
f_{z0} =\,&6{H}_0^2(j_0-q_0-2).\label{etdue}
\end{align}
\end{subequations}
These relations will be useful for cosmographic tests once suitable priors are fixed. Next we discuss how to achieve a deceleration-acceleration transition in the context of $f(\mathcal{R})$ gravity.

\section{$f(\mathcal{R})$ gravity and the cosmological deceleration-acceleration transition}

Ideally, we would like to integrate the modified Friedmann equations (\ref{primieroboth}) while taking into account Eqs.\ (\ref{Hinz0bi}) and (\ref{etunoboth}). However, we cannot solve Eqs.\ (\ref{primieroboth}) along with Eqs. (\ref{eq:rhocurv}) and (\ref{eq: wcurv}) directly, due to their complexity. We therefore assume a parameterized cosmological model, which can  depart  from $\Lambda$CDM at both low and high redshift. In particular, we find that a useful ansatz is a logarithmic correction associated to the dark energy term,
\begin{equation}\label{h}
{H}(z)={H}_0\sqrt{\Omega_m(1+z)^3+\log(\alpha+\beta z)}\,,
\end{equation}
where $\alpha$ and $\beta$ are constants. In order to get ${H}={H}_0$ at $z=0$, we require   $\alpha=\exp(1-\Omega_m)$, and to account for Eqs.\ (\ref{etunoboth}) we assume $\beta\in [0.01$,  $0.1]$. In so doing, we fix both $\alpha$ and $\beta$ in terms of mass and cosmographic coefficients. This prescription allows us to numerically reconstruct $f(z)$ in terms of cosmological data. Moreover, we will see later that these requirements for $\alpha$ and $\beta$ are compatible within $1\sigma$ errors with our observational results.

The ansatz for an $f(z)$ which results in the above $H(z)$ expression is
\begin{equation}\label{jhd}
f(z)= \tilde f_0+\frac{1}{1+z}+\tilde f_1 (1+z)^{\sigma_1}+\tilde f_2(1+z)^{\sigma_2}\,.
\end{equation}
This reproduces fairly well the numerical Friedmann equations up to $z\leq2$, and permits us to quantify the effects of $f(\mathcal R)$ gravity on $H$. We find a good agreement, with negligible departures from $z\ll1$ to $z\sim2$, for the parameter values $\tilde f_0\sim-10$, $\tilde f_1\sim 7$, $\tilde f_2\sim-3.7$, $\sigma_1=1$ and $\sigma_2=2$. These results are consistent with the cosmographic ranges of $f_0$ and $f_{z0}$.

As discussed above, in general relativistic dark energy cosmological models the universe switches from an earlier matter-dominated decelerating cosmological expansion to a later dark-energy dominated accelerating cosmological expansion. Recent improved cosmological data have now allowed for the first believable estimate of this deceleration-acceleration transition redshift, $z_{\mathrm{da}}$, \cite{ratra13} (for more recent developments see Ref \cite{Zhang}). We can test models by comparing theoretical predictions of the deceleration-acceleration transition redshift with observational data. The transition redshift, whose expression is formally given by
\begin{equation}\label{dn}
z_{\mathrm{da}}=\Big(\frac{1}{H}\frac{dH}{dz}\Big)^{-1}\Big|_{z=z_{\mathrm{da}}}-1\,,
\end{equation}
can be obtained by assuming $q=0$, corresponding to $\ddot a=0$.

In the $\Lambda$CDM model the acceleration parameter is
\begin{equation}\label{dkjnh}
q_{\Lambda}=\frac{3\Omega_m(1+z)^3}{2 + 2 \Omega_m z[3+z(3+z)]}\,,
\end{equation}
and the corresponding transition redshift is
\begin{equation}\label{trans}
z_{\mathrm{da},\Lambda}=\left[2\frac{(1-\Omega_m)}{\Omega_m}\right]^{1/3}-1\,.
\end{equation}
In the model we study,
\begin{equation}\label{quxz}
q_{f(\mathcal R)}=-1 + \frac{(1+z) \left[3\Omega_m(1+z)^2+{\beta}/(\alpha + \beta z)\right]}{ 2\left[\Omega_m(1+z)^3+ \ln(\alpha+\beta z)\right]}\,,
\end{equation}
and, in a first-order approximation around $z=0$, the transition redshift is
\begin{eqnarray}\label{lsdjkhk}
z_{\mathrm{da},f(\mathcal R)}=\quad\quad\quad\quad\quad\quad\quad\quad\quad\quad\quad\quad\quad\quad\quad\quad\quad\quad\quad\\
\frac{\beta\exp(\Omega_m-1)-2+3\Omega_m}{\left[3+2\beta\exp(\Omega_m-1)\right]\left[\beta\exp(\Omega_m-1)-2+3\Omega_m\right]}\,.\nonumber
\end{eqnarray}
In the following section we observationally constrain $z_{\mathrm{da},f(\mathcal R)}$ and compare its value with that from Eq. ($\ref{trans}$).

\section{Observational constraints on curvature dark energy parameterization}

In this section we use observational data to constrain the curvature dark energy parametrization discussed in the previous section. The observational data we use are type Ia supernova (SNIa) apparent magnitude versus redshift measurements and measurements of the Hubble parameter as a function of redshift. These data are shown in Fig.\ 1.

\begin{figure}[H]
\includegraphics[width=3.2in]{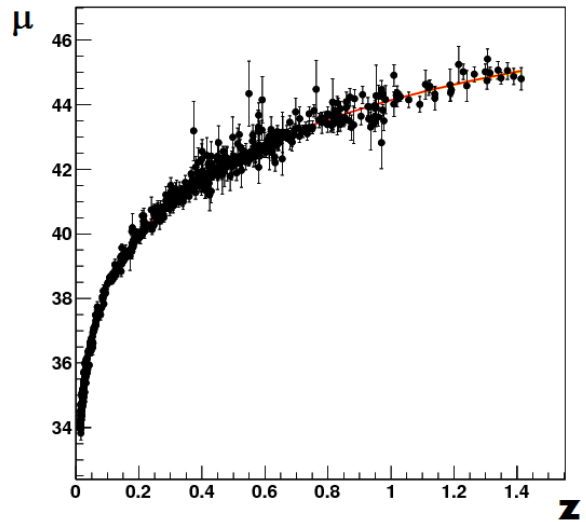}
\includegraphics[width=3.2in]{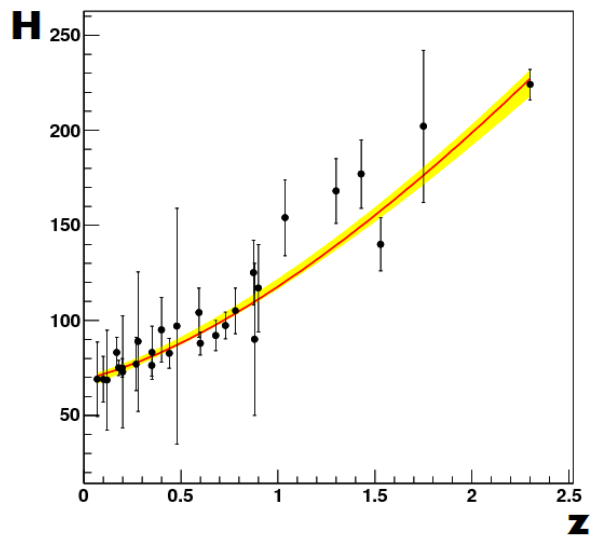}
{\small \caption{ Observational data with 1 $\sigma$ error bars and model predictions. Top panel show SNIa apparent magnitude data, bottom show $H(z)$ data (in the bottom panel $H$ is given in units of km s$^{-1}$ Mpc$^{-1}$). The red lines represent the best-fit (from the joint SNIa and $H(z)$ analysis) model prediction;  the width represents the 1 $\sigma$ uncertainty. }
\label{Fig:EoS}}
\end{figure}

\begin{figure}[hb]
\includegraphics[width=4in, height=3.5in]{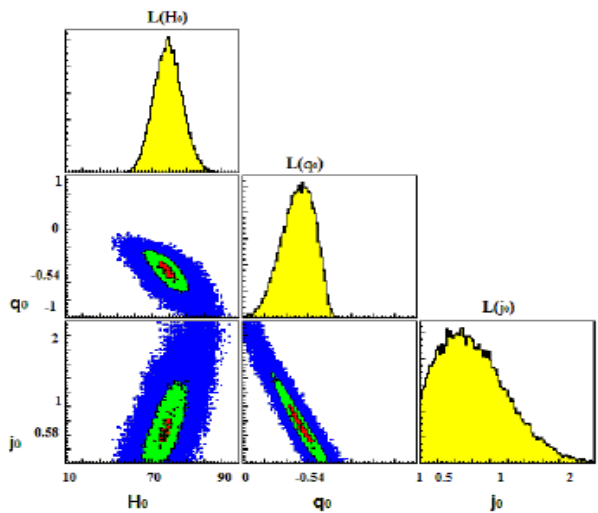}
\begin{center} \hspace{-0.2cm}
\caption{Two-dimensional marginalized constraint contour plots and one-dimensional marginalized probability density distribution functions for parameters of the cosmographic series computed using the Metropolis algorithm and Union 2.1 SNIa data.  The above contours correspond to allowing all cosmographic parameters to vary. $H_0$ values are in the units of km s$^{-1}$Mpc$^{-1}$.
}
\end{center}
\end{figure}

\begin{figure}
\begin{center}
\includegraphics[width=2.9in]{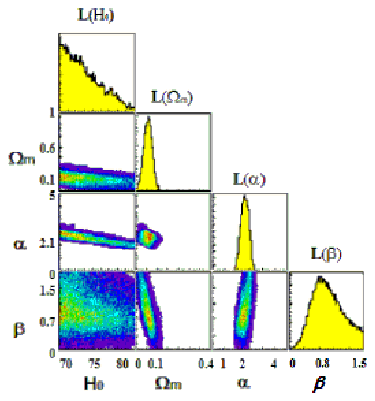}
\includegraphics[width=2.8in]{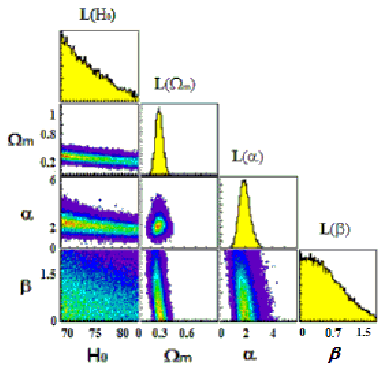}
\includegraphics[width=2.8in]{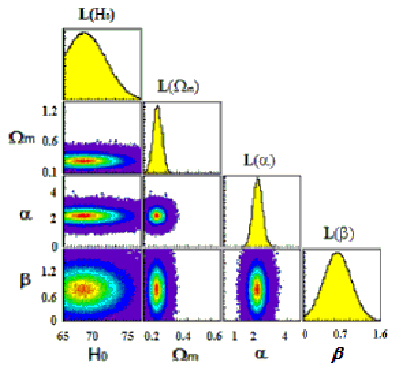}
\caption{Two-dimensional marginalized constraint contour plots and one-dimensional marginalized probability density distribution functions for free parameters  $H_0,\Omega_m,\alpha$ and $\beta$ of our model computed using the Metropolis algorithm. The three set of panels, from top to bottom, correspond to the results using: $1)$ the Union 2.1 SNIa data (top panels); $2)$ the $H(z)$ measurement compilation (middle panels); and, $3)$ a combined analysis using both the Union 2.1 SNIa data and the $H(z)$ measurement compilations (bottom panels). $H_0$ values are in the units of km s$^{-1}$Mpc$^{-1}$.
}
\end{center}
\end{figure}

The SNIa data we use are from the Union 2.1 compilation \cite{suzuky} of 580 supernovae up to redshift $z=1.414$; in our analyses here we account only for statistical uncertainties. In order to get SNIa data cosmological constraints we adopt the Monte Carlo technique based on the  Metropolis algorithm \cite{metro}, which reduces dependence on initial statistics. The likelihood function
\begin{equation} \label{funzionea}
	{\mathcal L}(\mathbf{p}) \propto \exp \left[-\chi^2(\mathbf{p})/2\right]\,,
\end{equation}
is maximized, and $\chi^2$ is therefore minimized. Here the free parameters are $\mathbf {p}=\left(H_0,q_0,j_0\right)$ and $\left(H_0,\Omega_m,\alpha,\beta\right)$ for the cosmographic parametrization and for the curvature dark energy model respectively. The $\chi^2$ function is
\begin{equation}
	\chi^{2}_{SN}(\mathbf{p}) =\sum_{k=1}^{580}\frac{\left[\mu_{D,k}^{\mathrm{th}}(\mathbf{p})-\mu_{D,k}^{\mathrm{obs}}\right]^{2}}
	{\sigma_{k}^{2}}\,,
\end{equation}
where the apparent magnitude
\begin{equation}\label{nagq}
\mu_{D.k}=25+5\log_{10}\left(\frac{d_{\mathcal L,k}}{1 Mpc}\right)
\end{equation}
for the $k^{\mathrm{th}}$ supernova. For cosmographic fits we use the expanded version of $d_\mathcal L$, whereas for fitting our model we employ $H$ of Eq. ($\ref{h}$).

The second data set we use are measurements of the Hubble parameter as a function of redshift $z$, $H(z)$, \cite{ultima1}. For our analyses we use the 28 $H(z)$ measurements given in Table 1 of Ref. \cite{ratra13}, with a highest redshift measurement at $z=2.3$. In this case
\begin{equation}\label{star}
\chi^{2}_{H}(\mathbf{p})=\sum_{l=1}^{28}\frac{H_{\mathrm{th,}l}^2(\mathbf{p})- H_{\mathrm{obs},l}^2}{\sigma_{H,l}^2}\,.
\end{equation}

In order to get interesting results we make a number of assumptions. First, we assume space curvature vanishes ($\Omega_{K}=0$) and ignore radiation. We also assume top hat priors for the parameters, listed in Table I. These are flat priors, non-zero inside and vanishing outside the listed range.

\newlength{\mywidth}
\setlength{\mywidth}{0.4\textwidth}
\begin{table}[H]
\caption{Priors imposed for the initial conditions on $f(z)$ parameters. The numerical values for $f_0$ and $f_{z0}$ are in units of $ H_0=100\ \mathrm{km\,s^{-1}\,Mpc^{-1}}$.}
\begin{center}
\begin{tabular}{c}
\begin{tabular*}{\mywidth}{c}
\hline\hline
\quad \quad \quad \quad Cosmological priors\\ \hline
\begin{tabular}{rcl}
\quad\quad\quad\quad\quad$0.5\quad <$ &$h$ & $< \quad 0.9$\\
$0.001 \quad <$&$\Omega_{\rm b}h^2$ & $< \quad 0.09$ \\
$\,-6 \quad <$&$ f_0 $&$< \quad  -1$ \\
$\,-3 \quad <$&$ f_{z0} $&$< \quad 0$ \\
$\,-4 \quad <$&$  q_0 $&$< \quad -0.1$ \\
$\,-5 \quad <$&$  j_0 $&$< \quad 5$
\end{tabular}\\  \hline
\hline
\end{tabular*}
\end{tabular}
\end{center}
\label{tab:priors}
\end{table}

In order to determine more restrictive constraints on the cosmographic parametrization, in this case we also perform analyses with a fixed value of $H_0$. More precisely the first value we use was determined from the \textit{Planck} data~\cite{Ade}, i.e., $H_0 = 67.11\ \mathrm{km\,s^{-1}\,Mpc^{-1}}$. The second $H_0$ value we use is that derived by fitting the low redshift, $z\leq 0.36$, Union 2.1 supernova apparent magnitude data to the first order luminosity distance, $d_\mathcal L = {z}/{H_0}$, resulting in $H_0 = 69.96_{-1.16}^{+1.12}\ \mathrm{km\,s^{-1}\,Mpc^{-1}}$. Both these values are consistent with other recent estimates. For instance, from a median  statistics analysis of 553 $H_0$ measurements Ref. \cite{G.chen} (for related work and results see Refs. \cite{gott,Colless}) find $H_0=68\pm2.8$ km s$^{-1}$ Mpc$^{-1}$. In our analysis here we ignore the small uncertainties in $H_0$.
\begin{table*}
\caption{{\small Best fit value and 1$\sigma$ error bars for each parameter of the cosmographic parametrization. We perform a no prior fit, in which all the cosmographic series is free to vary, a fit with $H_0$ fixed to the \textit{Planck} value, and a fit with $H_0$ fixed by using the first order luminosity distance $d_\mathcal L={z}/{H_0}$ with the Union 2.1 SNIa data for $z\leq0.36$, for the SNIa data in the second to fourth columns, and for the $H(z)$ data in the last three columns. $H_0$ values are in units of km s$^{-1}$ Mpc$^{-1}$.}}

\begin{tabular}{c c c c c c c}
\hline
\hline

fit 			& SNIa, free & SNIa, \textit{Planck} $H_0$ & SNIa, our $H_0$ & $H(z)$, free & $H(z)$, \textit{Planck} $H_0$ & $H(z)$, our $H_0$  \\ [0.8ex]
\hline
$p$ value		&  {\small $0.6899$ }	
                &  {\small $0.2381$ }
                &  {\small $0.6896$ }
                &  {\small $0.0691$ }
    			&  {\small $0.0268$ } 	
                &  {\small $0.1206$ } \\[0.8ex]

\hline
{\small$H_0$}		& ~{\small $69.97$}{\tiny${}_{-0.41}^{+0.42}$}~	
                            & ~{\small $67.11$}{$\,\mathrm{fixed}$}~
            			    & ~{\small $69.96$}{$\,\mathrm{fixed}$}~	
                            & ~{\small $66.38$}{\tiny${}_{-1.04}^{+2.36}$}~	
                            & ~{\small $67.11$}{$\,\mathrm{fixed}$}~
            			    & ~{\small $69.96$}{$\,\mathrm{fixed}$}~ \\[0.8ex]

{\small$q_0$}		& ~{\small $-0.5422$}{\tiny${}_{-0.0826}^{+0.0718}$}~	
                    & ~{\small $-0.0732$}{\tiny${}_{-0.0529}^{+0.0538}$}~
		            & ~{\small $-0.5319$}{\tiny${}_{-0.0465}^{+0.0520}$}~	
                    & ~{\small $-2.9412$}{\tiny${}_{-0.0426}^{+0.0922}$}~
                    & ~{\small $-6.8930$}{\tiny${}_{-0.0749}^{+0.1628}$}~
			        & ~{\small $-2.9213$}{\tiny${}_{-0.2589}^{+0.2688}$}~ \\[0.8ex]

{\small$j_0$}		& ~{\small $ 0.5768$}{\tiny${}_{-0.3528}^{+0.4478}$}~	
                    & ~{\small $-0.8957$}{\tiny${}_{-0.1828}^{+0.1948}$}~
        			& ~{\small $0.5112$}{\tiny${}_{-0.3035}^{+0.2831}$}~	& ~{\small $-0.955$}{\tiny${}_{-0.175}^{+0.228}$}~
		          	& ~{\small $0.1249$}{\tiny${}_{-0.8318}^{+1.6899}$}~ 	
                    & ~{\small $-3.9040$}{\tiny${}_{-2.2510}^{+3.4030}$}~\\[0.8ex]

\hline \hline

\end{tabular}



\label{tab:dL}
\end{table*}

\begin{table*}
\caption{{\small Best fit value and 1$\sigma$ error bars for each parameter of our model. We perform a fit by using SNIa data with no priors imposed \emph{a priori} (column two), using $H(z)$ data only (column three), and by using the combined SNIa and $H(z)$ data together (column four). For these three fits, the parameters $H_0, \Omega_m, \alpha$ and $\beta$ are free to vary. The transition redshifts have been evaluated by means of Eq. ($\ref{lsdjkhk}$), and errors are estimated through standard logarithmic propagation. In so doing, we used the estimated values of $\Omega_m$ and $\beta$ along with the condition $\alpha=\exp(1-\Omega_m)$. $H_0$ values are  in units of km s$^{-1}$ Mpc$^{-1}$.}}

\begin{tabular}{c c c c}
\hline
\hline

fit 			& SNIa, free &  $H(z)$, free & Combined \\ [0.8ex]
\hline
$p$ value		&  {\small $0.6919$ }
                &  {\small $0.9604$ }
                &  {\small $0.6885$ } \\[0.8ex]

\hline
{\small$ H_0$}		& ~{\small $68.07$}{\tiny${}_{-2.20}^{+3.33}$}~	
                            & ~{\small $68.49$}{\tiny${}_{-2.50}^{+3.39}$}~
                            & ~{\small $67.94$}{\tiny${}_{-1.82}^{+2.20}$}~\\[0.8ex]

{\small$\Omega_m$}		& ~{\small $0.2142$}{\tiny${}_{-0.0413}^{+0.0386}$}~	
                        & ~{\small $0.2718$}{\tiny${}_{-0.0326}^{+0.0335}$}~
                        & ~{\small $0.2316$}{\tiny${}_{-0.0391}^{+0.0391}$}~\\[0.8ex]

{\small$\alpha$}		& ~{\small $2.3011$}{\tiny${}_{-0.1690}^{+0.1648}$}~	
                        & ~{\small $1.9660$}{\tiny${}_{-0.1991}^{+0.2307}$}~
                        & ~{\small $2.339$}{\tiny${}_{-0.1853}^{+0.1879}$}~ \\[0.8ex]

{\small$\beta$}		& ~{\small $0.7599$}{\tiny${}_{-0.4712}^{+0.6429}$}~	
           			& ~{\small $0.2846$}{\tiny${}_{-0.5372}^{+0.7734}$}~	
                    & ~{\small $0.6101$}{\tiny${}_{-0.4351}^{+0.4388}$}~\\[0.8ex]
		
{\small$z_{\mathrm{da}}$}		& ~{\small $0.8596$}{\tiny${}_{-0.2722}^{+0.2886}$}~	
                            & ~{\small $0.6320$}{\tiny${}_{-0.1403}^{+0.1605}$}~
                            & ~{\small $0.7679$}{\tiny${}_{-0.1829}^{+0.1831}$}~\\[0.8ex]

\hline \hline

\end{tabular}



\label{tab:dL2}
\end{table*}

To derive constraints on the parameters of our model, standard procedures with no priors were used. In particular, we consider three tests. The first uses supernovae, the second is with $H(z)$ measurements, and the third combines supernovae measurements with $H(z)$ data (by minimizing $\chi^2_{\mathrm{tot}}=\chi^2_{SN}+\chi^2_{H}$).

Supposing the validity of the null hypothesis for each fit, we report the corresponding $p$-values---the probability that a result obtained by a single fit is observed---representing a qualitative measure of the likelihood for a certain outcome. Our results were obtained by using the publicly available code ROOT \cite{root} and the bayesian toolkit BAT \cite{bat}.

Figures\ 2 and 3 show the resulting two-dimensional constraint contours and corresponding one-dimensional probability density distribution function for the model parameters. In the contour plots, different colours indicate the $68\%$, $95\%$ and $99\%$ confidence level regions. From these figures we see that the cosmological parameters of the cosmographic parametrization and of the model we consider are quite tightly constrained by the $68$\% confidence level contours.

We summarize our numerical results in Tables II and III. Our outcomes seem to favor values of the Hubble constant consistent with other estimates \cite{G.chen, gott, Colless}. The \textit{Planck} priors on $H_0$ lead to rather low $p$ values (low goodness of fit) whereas our prior on $H_0$, derived from fitting supernovae in the redshift range $z\in[0,0.36]$, leads to higher $p$, statistically favored, best fits

In addition, we find that the inferred limits on $f_0$ and $f_{z0}$,  used in Sec. II for numerically solving the modified Friedmann equations, are compatible with our experimental results. In Table III we also list constraints on $\alpha$ and $\beta$ which are in agreement with theoretical predictions. In Fig.\ 4  we plot the acceleration parameter $q$ using the best-fit parameter value reported in the fourth column of Table III. The acceleration parameter changes sign at a transition redshift around $z_{\mathrm da}\sim 0.8$. This is in agreement with the transition redshift measured in Ref. \cite{ratra13}.

In order to infer numerical values for the transition redshift, we use Eq.\ ($\ref{lsdjkhk}$) and the estimated value of $\Omega_m$ and $\beta$. The obtained transition redshifts are listed in the last line of Table III. These are in the range $z_\mathrm{da}\in[0.57,0.97]$. We determine errors on $z_\mathrm{da}$ by standard logarithmic error propagation. These results are not incompatible with the observed value \cite{ratra13}.

These results on $z_\mathrm{da}$ are also compatible with the $\Lambda$CDM model prediction, i.e.\ Eq.\ ($\ref{trans}$), which  leads to a transition redshift within the interval $z_{\mathrm{da},\Lambda}\in[0.67,1]$. In addition, we conclude that our transition redshifts are compatible with the priors of Table I. Future and more accurate $z_{\mathrm da}$ measurements will improve the accuracy and will permit us to better distinguish any significant deviation from the $\Lambda$CDM model.

\begin{figure}
\includegraphics[width=3.2in]{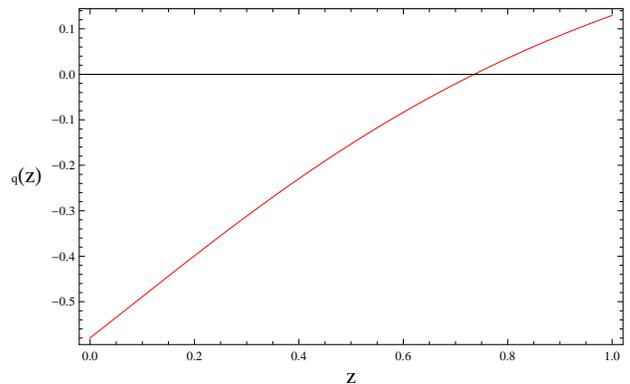}
\caption{Acceleration parameter as a function of redshift, Eq.\ (24), for our model, using best-fit parameter values from the combined SNIa and $H(z)$ data analysis reported in the last column of Table III. The acceleration parameter changes sign at a transition redshift $z_{\mathrm{da}}\sim 0.8$. More accurate results for the transition redshift are reported in Table III.}
\label{Fig:EoS}
\end{figure}

\section{A possible physical interpretation of curvature dark energy}

In Sec.\ IV we showed that  extensions  of Einstein's gravity, in particular  $f(\mathcal{R})$ gravity, may lead to  logarithmic corrections to the conventional Hubble parameter and, in particular, a dark energy  term of the form $\Omega_{DE}=\ln(\alpha+\beta z)$. Here we discuss the physical meaning of such a correction, interpreting $\alpha$ and $\beta$, i.e.\ the free parameters which enter Eq.\ ($\ref{h}$), in terms of $f(\mathcal R)$.

To this end we first note that one can expand $f(\mathcal{R})$ in terms of the Ricci scalar evaluated at the present epoch, $\mathcal{R}_0$. This expansion turns out to be compatible with current cosmographic requirements, as shown in Secs.\ II and III. We also assume that the gravitational constant $G$ is time independent now. In addition, at $\mathcal{R}=\mathcal{R}_0$ the second derivative of $f(\mathcal{R})$ should be negligibly small, so that the Solar System constraints are satisfied (see for example \cite{capozziello}). Thus, in the Taylor expansion of $f(\mathcal{R})$
\begin{eqnarray}\label{massa2}
f(\mathcal{R})=&& f(\mathcal{R}_0)+f'(\mathcal{R}_0)\left(\mathcal{R}-\mathcal{R}_0\right)+\nonumber\\
&&+\frac{1}{2}f''(\mathcal{R}_0)\left(\mathcal{R}-\mathcal{R}_0\right)^2+\nonumber\\
&&+\frac{1}{6}f'''(\mathcal{R}_0)\left(\mathcal{R}-\mathcal{R}_0\right)^3+\ldots\,,
\end{eqnarray}
the above mentioned constraints require
\begin{subequations}\label{constraint1}
\begin{align}
f'(\mathcal{R}_0) &= 1\,,\\
f''(\mathcal{R}_0) &= 0\,,
\end{align}
\end{subequations}
allowing for the observational viability of the model at the present epoch. Using \cite{grandecapozziello}
\begin{equation}
\label{zuzzu1}
f'''(\mathcal{R})=\, \frac{f_{3z}}{\mathcal{R}_z^3} - \frac{f_z\, \mathcal{R}_{3z}+3f_{2z}\,
\mathcal{R}_{2z}}{\mathcal{R}_z^4}+\frac{3f_z\, \mathcal{R}_{2z}^2}{\mathcal{R}_z^5}\,,
\end{equation}
with
\begin{eqnarray}
\begin{split}
\frac{\mathcal{R}_{3z0}}{6}=\,&3\mathcal{H}_{2z0}^2+\\
&\mathcal{H}_{z0}\left(-3\mathcal{H}_{2z0}+4\mathcal{H}_{3z0}\right)
+{H}_0(-\mathcal{H}_{3z0}+\mathcal{H}_{4z0})\,,
\end{split}
\end{eqnarray}
and Eq.\ (\ref{eq:rhocurv}), $\rho_{\mathrm{curv0}} = \left[{f(\mathcal{R}_0) - \mathcal{R}_0}\right]/{2}$, we have
\begin{equation}\label{vtgu}
f_0  =6 {H}_0^2 (1 - \Omega_m) + \mathcal{R}_0\,.
\end{equation}

As a consequence of the aforementioned constraints, using Eq.\ ($\ref{zuzzu1}$) we can express $\alpha$ and $\beta$ in terms of $f_0$ and $f_{0}^{'''}$, showing that these two parameters depend on $f(\mathcal{R})$ and its derivatives with respect to $\mathcal{R}$, around $\mathcal{R}=\mathcal{R}_0$. One relation between $\alpha$ and $\beta$ in terms of $f_0$ is
\begin{eqnarray}\label{betax}
\beta=&&\alpha\Omega_m\frac{f_0 - 3 {H}_0^2 (1 - 2 \Omega_m)}{3{H}_0^2}\nonumber\\
&&+ \log\alpha\frac{f_0 + 6 {H}_0^2 (1 + \Omega_m)}{3{H}_0^2}.
\end{eqnarray}
Analogously, one may infer another relation between $\alpha$ and $\beta$ in terms of the third derivatives, by using Eq.\ (\ref{zuzzu1}),
$\alpha=\alpha\left(f_0,f'''(\mathcal{R}_0)\right)$; the explicit form is not important for our purposes. Since the physical significance of the leading term in  the Taylor expansion of $f(\mathcal{R})$ is well established \cite{furth}, the above relations allow for an understanding of the physical significance of $\alpha$ and $\beta$.

For our purposes, the zero order $f_0$ term corresponds to an initial value cosmological constant  $(1-\Omega_m)$. This means that the coincidence problem can be  reinterpreted in $f(\mathcal{R})$ gravity as the choice of initial  conditions for the corresponding curvature dark energy term. Moreover, by taking into account the parameterized post-Newtonian (PPN) approximation, up to the second order in  $f(\mathcal{R})$, and considering the first two parameters of the Eddington parametrization, $\beta^{(PPN)}$ and $\gamma^{(PPN)}$, we can write \cite{troi}
\begin{equation}\label{betaR}
\beta^{PPN}_{\mathcal R}-1\,=\,\frac{f'(\mathcal{R}) f''(\mathcal{R})}{8 f'(\mathcal{R})+12f''(\mathcal{R})^2}\frac{d\gamma^{PPN}_{\mathcal R}}{d\mathcal R}\,,
\end{equation}
and
\begin{equation}\label{gammaR}
\gamma^{PPN}_{\mathcal R}-1\,=-\,\frac{f''(\mathcal{R})^2}{f'(\mathcal{R})+2f''(\mathcal{R})^2}\,.
\end{equation}
 Solar system constraints on $\beta^{PPN}$ and $\gamma^{PPN}$ are not violated because we assume $f''(\mathcal{R}_0)=0$, and so general relativity, i.e. $\beta^{(PPN)}=\gamma^{(PPN)}=1$, is locally valid.

As a consequence, it seems that gravitational corrections due to $f(\mathcal{R})$ gravity become significant at the third order of the expansion. In other words,  curvature dark energy, inferred from  $f(\mathcal{R})$ and compatible with current cosmographic bounds, gives  contributions at the third order of $f(\mathcal R)$ expansions. Rephrasing it, the corresponding cosmological model reduces to $\Lambda$CDM when  corrections only up to second order are included.

\section{Conclusion}

We have numerically analyzed a class of $f\left(\mathcal{R}\right)$ gravity models which reduce to $\Lambda$CDM at $z\simeq0$.  Deviations emerge at third order in the $f(\mathcal R)$ Taylor expansion.   As present epoch constraints  we adopt the cosmographic series, i.e. the series of measurable coefficients derived by expanding the luminosity distance and comparing it with data. We therefore inferred cosmographic bounds on the test function $f(z)$ which reproduces the observed low redshift cosmological behavior.

Since cosmography allows for a determination of model independent constraints on $f(z)$ and derivatives, we used a Taylor expansion of $f(z)$ in terms of $a(t)={1}/{(1+z)}$, which fairly well approximates the Friedmann equations in the range $z\leq 2$. We found good agreement, with small departures at $z\ll1$ and $z\sim2$, for the range of parameters $\tilde f_0\sim-10$, $\tilde f_1\sim 7$, $\tilde f_2\sim-3.7$, $\sigma_1=1$ and $\sigma_2=2$, which are compatible with the initial conditions defined by cosmography.

Such departures lead to possible logarithmic corrections of the conventional Hubble rate, showing an evolving dark energy term different from the cosmological constant. We demonstrated that this model has a transition redshift in a  range compatible  with measurements \cite{ratra13}. To this end, cosmological constraints on the model were determined using a Monte Carlo approach based on the Metropolis algorithm. Our model passes all the cosmological tests, showing that the obtained curvature dark energy is compatible with observations. We implemented different priors on the fitting parameters, and in particular, we fixed $H_0$ to the \textit{Planck} value first and then to a numerical value obtained by fitting the first-order luminosity distance $d_\mathcal L$ to the supernova data in the interval $z\in[0,0.36]$. In general,  results seem to indicate slightly less negative acceleration parameters with non-conclusive results on the variation of acceleration, namely the jerk parameter.

Using these results we provided a self consistent explanation of the free parameters of the model, showing that they could be related to  the terms of the  Taylor series of $f(\mathcal{R})$. In doing so, by comparing our results with PPN approximations, we found that $\alpha$ and $\beta$ could be related to third-order PPN parameters. Future investigations will be devoted to better constraining the logarithmic correction due to $f(\mathcal R)$.

\acknowledgements

SC and OL thank  Manuel Scinta for useful discussions and support in the Monte Carlo analysis. Part of this work was done at the South National Laboratories of Nuclear Physics at the University of Catania. SC is  supported by INFN (iniziative specifiche NA12, OG51). OL is supported by the European PONa3 00038F1 KM3NET (INFN) Project. OF and BR are supported in part by DOE grant DEFG03-99EP41093 and NSF grant AST-1109275.



%
%




\begin{thebibliography}{}


\bibitem{Peebles84}
P. J. E. Peebles, Astrophys J. {\bf 284}, 439 (1984).

\bibitem{Sami}
e.g., M. Sami and R. Myrzakulov, arXiv:1309.4188 [hep-th]; M. J. Mortonson, D. H. Weinberg, and M. White, arXiv:1401.0046 [astro-ph.CO].

\bibitem{lambdaobs}
e.g., P. J. E. Peebles, and B. Ratra, Rev. Mod. Phys {\bf 75}, 559 (2003);  D. H. Weinberg, \textit{et. al.}, arXiv:1306.0913 [astro-ph.CO].

\bibitem{Sola}
e.g., P. Bin\'{e}tury, Astron. Astrophys Rev. {\bf 21}, 67  (2013); C.P. Burgess, arXiv:1309.4133 [hep-th].


\bibitem{darkenergy}
S. Nojiri and S. D. Odintsov, Phys. Rep. {\bf 505}, 59 (2011).

\bibitem{Podariu}
e.g., S. Podariu and B. Ratra, Astrophys J. {\bf 532}, 109 (2000).

\bibitem{PeeblesandRatra}
 P. J. E. Peebles and B. Ratra, Astrophys. J. Lett. {\bf 325} L17 (1988); B. Ratra and P. J. E. Peebles, Phys. Rev. D, {\bf 37}, 3406 (1988).


\bibitem{mico}
O. Luongo and H. Quevedo, Astrophys. Space Sci. {\bf 338}, 345 (2011); O. Luongo and H. Quevedo, Int. J. Mod. Phys. D, {\bf 23}, 1450012, (2014); also see K.-H. Chae, G. Chen, D.-W. Lee, and B. Ratra, Astrophys. J. Lett. {\bf 607}, L71 (2004); L. Samushia and B. Ratra, Astrophys. J. {\bf 714}, 1347 (2010); Y. Wang and S. Wang, Phys. Rev. D {\bf 88}, 043522 (2013). R. F. L. Holanda, J. W. C. Silva, and F. Dahia, Class. Quant. Grav. {\bf 30}, 205003 (2013); X. Wang, X-L Meng, Y. F. Huang, and T. J. Zhang, Res. Astron. Astrophys. {\bf 13}, 1013 (2013); J. Bielefeld, W. L. K. Wu, R. R. Caldwell, and O. Dor\'{e}, Phys Rev D {\bf 88}, 103004 (2013); S. Thakur and A. A. Sen, arXiv.1305.6447 [astro-ph.CO]; E. L. D. Perico, J. A. S. Lima, S. Basilakos, and J. Sol\`{a} Phys. Rev. D {\bf 88}, 063531 (2013); S. Crandall and B. Ratra, arXiv: 1311.0840 [astro-ph.CO]; A. Pavlov, O. Farooq, and B. Ratra, arXiv:1312.5285 [astro-ph.CO].

\bibitem{notes}
C. Rubano and P. Scudellaro, Gen. Rel. Grav. {\bf 34}, 1931 (2001); N. Straumann, Mod. Phys. Lett. A {\bf 21}, 1083 (2006); E. V. Linder, arXiv:1009.1411 [gr-qc].

\bibitem{coppa4}
A. V. Astashenok and S. D. Odintsov, Phys. Lett. B {\bf 718}, 1194 (2013); K. Bamba, S. Nojiri, and S. D. Odintsov, Proc. 7th Math. Phys. Meet., Belgrade, Serbia (2012).

\bibitem{coppa45}
S. Nojiri, and S. D. Odinstov, arXiv:0807.0685 [hep-th]; also see R. P. Woodard, arXiv:1401.0254 [astro-ph.CO].

\bibitem{curvatura}
e.g., A. A. Starobinsky, JETP Lett. {\bf 86}, 157 (2007); G. J. Olmo, Phys. Rev. D {\bf 72}, 083505 (2005); S. Tsujikawa, Phys. Rev. D {\bf 77}, 023507 (2008); G. Cognola, \textit{et. al.}, Phys. Rev. D {\bf 79}, 044001, (2009).

\bibitem{geometric}
S. Capozziello and M. De Laurentis, Phys. Rept. {\bf 509}, 167 (2011).

\bibitem{beng}
A. Aviles, A. Bravetti, S. Capozziello, and O. Luongo, Phys. Rev. D {\bf 87}, 044012 (2013).


\bibitem{ratra13}
O. Farooq and B. Ratra, Astrophys. J. Lett. {\bf 766}, L7 (2013).

\bibitem{cosmography1}
M. Visser, Gen. Rel. Grav. {\bf 37}, 1541 (2005); S. Weinberg, \emph{Cosmology}, Oxford Univ. Press, Oxford (2008); C. Cattoen and M. Visser, Phys. Rev. D {\bf 78}, 063501 (2008).

\bibitem{cosmography2}
M. Visser, Class. Quant. Grav. {\bf 21}, 2603 (2004); S. Capozziello and V. Salzano, Adv. Astron. {\bf 2009}, 217420 (2009).

\bibitem{cosmography3}
M. Demianski, E. Piedipalumbo, C. Rubano, and P. Scudellaro, Mon. Not. Roy. Astr. Soc. {\bf 426}, 1396 (2012); M. Arabsalmani and V. Sahni, Phys. Rev. D {\bf 83}, 043501 (2011).

\bibitem{cosmography4}
A. R. Neben and M. S. Turner,  Astrophys. J. {\bf 769}, 133, (2013); A. Aviles, C. Gruber, O. Luongo, and H. Quevedo, arXiv:1301.4044 [gr-qc]; M. Visser and C. Cattoen, Class. Quant. Grav. {\bf 24}, 5985 (2007).

\bibitem{suzuky}
N. Suzuki, {\it et al.} (The Supernova Cosmology Project), Astrophys. J. {\bf 746}, 85 (2012).

\bibitem{ultima1}
J. Simon, L. Verde, and R. Jimenez, Phys. Rev. D {\bf 71}, 123001 (2005); D. Stern, \emph{et al.}, J. Cosmol. Astropart. Phys. {\bf 1002}, 008 (2010); M. Moresco, \emph{et al.}, J. Cosmol. Astropart. Phys. {\bf 1208}, 006 (2012); C. Blake, \emph{et al.}, Mon. Not. Roy. Astr. Soc. {\bf 425}, 405 (2012); C. H. Chuang and Y. Wang, Mon. Not. Roy. Astr. Soc. {\bf 426}, 226 (2012); C. Zhang, \emph{et al.},   arXiv:1207.4541 [astro-ph.CO]; N. G. Busca, \emph{et al.}, Astron. Astrophys. {\bf552}, A96 (2013).

\bibitem{metro}
N. Metropolis, \textit{et al.}, J. Chem. Phys. {\bf 21}, 1087 (1953); H. M$\ddot{\mathrm{u}}$ller-Krumbhaar and K. Binder, J. Stat. Phys. {\bf 8}, 1 (1973).

\bibitem{root}
http://root.cern.ch/drupal/

\bibitem{bat}
https://www.mppmu.mpg.de/bat/

\bibitem{cosm}
A. Aviles, C. Gruber, O. Luongo, and H. Quevedo, Phys. Rev. D {\bf 86}, 123516 (2012).

\bibitem{lymy}
S. Capozziello, M. de Laurentis, and V. Faraoni, Open Astron. J. {\bf 3}, 49 (2009).

\bibitem{turnercosmografia}
e.g., R. D. Blandford, \textit{et al.}, arXiv:astro-ph/0408279; A. R. Neben and M. S. Turner, Astrophys. J. {\bf 769}, 133 (2013). Z.-X. Zhai, \textit{et al.}, Phys. Lett. B {\bf 727}, 8 (2013).

\bibitem{Ade}
P. A. R, Ade, \textit{et al.}, arXiv:1303.5076 [astro.ph.CO]; for an early indication see S. Podariu, \textit{et al.}, Astrophys. J. 559, {\bf 9}, (2001).

\bibitem{Anatoly2}
A. Pavlov, S. Westmoreland, K. Saaidi, and B. Ratra, Phys. Rev. D {\bf 88}, 123513 (2013); O. Farooq, D. Mania, and B. Ratra, arXiv.1308.0834 [astro.ph.CO], and references therein.

\bibitem{S.Podariu}
e.g., S. Podariu, P. Nugent, and B. Ratra, Astrophys J. \textbf{553}, 39 (2001); L. Samushia, \textit{et al.}, Mon. Not. Roy. Astron. Soc. \textbf{410}, 1993 (2011); B. Sartoris, S. Borgani, P. Rosati, and J. Weller, Mon. Not. Roy. Astron. Soc. \textbf{423}, 2503 (2012); T. Basse, O. E. Bjaelde, S. Hannestad, and Y. Y. Y. Wong, arXiv:1205.0548 [astro-ph.CO]; A Pavlov, L. Samushia, and B. Ratra, Astrophys J. \textbf{760}, 19 (2012); S. A. Appleby and E. Linder, Phys. Rev. D. \textbf{87}, 023532 (2013); M. Arabsalmani, V. Sahni, and T. D. Saini, Phys. Rev . D. \textbf{87}, 083001 (2013).

\bibitem{at}
J. V. Cunha, Phys. Rev. D {\bf 79}, 047301 (2009); J. V. Cunha, Mon. Not. Roy. Astron. Soc. {\bf 390}, 210 (2008).

\bibitem{att}
M. J. Mortonson, W. Hu, and D. Huterer, Phys. Rev. D, {\bf 80}, 067301 (2009); F. Y. Wang and Z. G. Dai, Mon. Not. Roy. Astron. Soc., {\bf 368}, 371 (2006).

\bibitem{vs2}
V. Sahni, T.D. Saini, A. A. Starobinsky, and U. Alam JETP Lett. {\bf 77}, 201 (2003); O. Luongo, Mod. Phys. Lett. A {\bf 26}, 20, 1459, (2011); O. Luongo, Mod. Phys. Lett. A {\bf 28}, 1350080 (2013).

\bibitem{salo}
S. Capozziello, V. F. Cardone, and V. Salzano, Phys. Rev. D {\bf 78}, 063504 (2008).

\bibitem{Zhang}
M. J. Zhang, \textit{et al.}, Phys. Rev. D {\bf 88}, 063534 (2013); Z.-X. Zhai, \textit{et al.}, Phys. Lett. B {\bf 727}, 8 (2013); O. Farooq, S. Crandall, and B. Ratra, Phys. Lett. B {\bf 726}, 72 (2013); O. Akarsu, T. Dereli, S. Kumar, and L. Xu, Eur. Phys. J. Plus {\bf 129}, 22 (2014);  V. Poitras, arXiv:1307.6172 [astro-ph.CO]; J. Lu, \textit{et al.}, Int. J. Mod. Phys. D, {\bf 22}, 1350059 (2013); L. P. Chimento and M. G. Richarte, Eur. Phy. J. C {\bf73}, 2497 (2013); Q. Gao and Y. Gong, arXiv:1308.5627 [astro-ph.CO]; C. Gruber and O. Luongo, arXiv:1309.3215 [gr-qc]; K. Bamba \textit{et al.}, arXiv:1309.6413 [hep-th]; V.C. Busti, R. F. L. Holanda, and C. Clarkson, J. Cosmol. Astropart. Phys. {\bf 1311}, 020 (2013); P. C. Ferreira, D. Pav\'{o}n, and J. C. Carvalho, Phys Rev. D {\bf 88}, 083503 (2013).

\bibitem{G.chen}
G. Chen and B. Ratra, Publ. Astron. Soc. Pacific \textbf{123}, 1127 (2011).

\bibitem{gott}
J. R. Gott, M. S. Vogeley, S. Podariu, and B. Ratra, Astrophys J. \textbf{549}, 1 (2001); G. Chen, J. R. Gott, and B. Ratra, Publ. Astron. Soc. Pacific \textbf{115}, 1269 (2003); E. Calabrese, M. Archidiacono, A. Melchiorri, and B. Ratra, Phy Rev D \textbf{86}, 043520 (2012).

\bibitem{Colless}
M.Colless, F. Beutler, and C Blake, arXiv:1211.2570 [astro-ph.CO]; G. Hinshaw, \textit{et al.}, Astrophys. J. Supp. \textbf{208}, 19 (2013); P. A. R, Ade, \textit{et al.}, arXiv:1303.5076 [astro.ph.CO]; G. Efstathiou, arXiv:1311.3461 [astro-ph.CO].

\bibitem{capozziello}
S. Capozziello, V. F. Cardone, and A. Troisi, Phys. Rev. D, {\bf 71}, 043503 (2005).

\bibitem{grandecapozziello}
M. Bouhmadi-Lopez, S. Capozziello, and V. F. Cardone, Phys. Rev. D {\bf 82}, 103526 (2010).

\bibitem{furth}
K. Bamba, S. Capozziello, S. Nojiri, and S. D. Odintsov, Astrophys. Space Sci. {\bf 342}, 155 (2012).

\bibitem{troi}
S. Capozziello and A. Troisi, Phys. Rev. D, {\bf 72}, 044022, (2005).


\end{thebibliography}
\end{document}